# From Administrative Chaos to Analytical Cohorts: A Three-Stage Normalisation Pipeline for Longitudinal University Administrative Records

Hugo Roger Paz
PhD Professor and Researcher Faculty of Exact Sciences and Technology National University of Tucumán
Email: hpaz@herrera.unt.edu.ar

**ABSTRACT**

The growing use of longitudinal university administrative records in data-driven decision-making often overlooks a mundane but critical layer: how raw, historically inconsistent data are normalised before any modelling takes place. This article presents a three-stage normalisation pipeline for a census-like dataset of 24,133 engineering students from a Latin American public university, covering more than four decades of enrolment and progression records (1980–2019), enriched with demographic and secondary-school background information.

The pipeline comprises: (i) **N1 CENSAL**, which harmonises demographic fields into a single person-level census layer; (ii) **N1b IDENTITY RESOLUTION**, which detects and consolidates duplicate or inconsistent identifiers into a canonical person identifier while preserving a full audit trail; and (iii) **N1c GEO & SECONDARY-SCHOOL NORMALISATION**, which builds reference tables for countries, provinces, localities and secondary schools, classifies school type (state national, state provincial, private secular, private religious), and explicitly flags irrecoverable cases as *DATA_MISSING*. Across iterations, the pipeline preserves 100% of students, achieves full geocoding of place of birth and residence, and yields valid secondary-school type classification for 56.6% of the population, with 43.4% correctly identified as structurally missing due to legacy enrolment practices rather than stochastic non-response.

A forensic missingness analysis combines chi-square tests and logistic regression to show that missing secondary-school information is highly predictable from decade of entry, degree programme and geography, indicating a structural, historically induced missingness mechanism rather than random or student-driven omission. The article contributes: (a) a transparent, reproducible normalisation pipeline tailored to higher-education administrative records, (b) an explicit framework for treating structurally missing secondary-school information without speculative imputation, and (c) practical guidance on how to define analytically coherent cohorts (full population vs. secondary-school-informed subcohorts) for downstream learning analytics, causal inference and policy evaluation.

**KEYWORDS**

longitudinal university administrative records, engineering education, data cleaning and normalisation, secondary-school background, missing data mechanisms, higher education analytics, data quality audit.

# 1. INTRODUCTION

Longitudinal administrative records are one of the few data sources capable of showing how real students actually move through higher education systems over decades rather than semesters. In engineering programmes, where time-to-degree routinely departs from formal curriculum duration, these records are essential to understand who persists, who leaves, and under what structural conditions (Herzog, 2006; Peugh & Enders, 2004). Yet, despite their potential, administrative datasets are often treated as if they were neutral and complete, with very limited attention to how missing values, historical changes in information systems, and inconsistent coding practices shape any downstream analysis (Cheema, 2014; Enders, 2010).

The last four decades have seen the massification of higher education worldwide, including in public systems with open or quasi-open access, strong social demand for engineering degrees, and constrained institutional resources (Altbach et al., 2010). In such contexts, engineering programmes frequently carry the dual mandate of expanding access while maintaining professional standards. This tension is usually studied through survey data, small cohort studies, or cross-sectional indicators of efficiency. Large-scale longitudinal administrative data are less visible in the literature, even though they provide the natural substrate to reconstruct complete student trajectories, model time-to-event processes, and examine how institutional rules and policies interact with individual biographies over long periods (Herzog, 2006).

Working with these data, however, is technically and conceptually demanding. First, the very notion of a "student" is not stable in real institutional systems: individuals can appear under multiple identifiers, change degree programmes, re-enter after long absences, or coexist in parallel legacy and modern information systems. Second, key censal variables—such as secondary school background, place of origin, or employment status—are often introduced gradually as information systems evolve, producing structural rather than random missingness (Little & Rubin, 2019). Third, naïve handling of missing data (e.g., listwise deletion, ad-hoc recoding) can distort estimates of retention, completion, and equity, especially when missingness is related to time, geography, or degree type (Cheema, 2014; Peugh & Enders, 2004).

Research on missing data in education and the social sciences has repeatedly shown that the mechanism behind missingness—missing completely at random, missing at random, or missing not at random—is more important than the percentage of missing values itself (Enders, 2010; Little & Rubin, 2019). Methodological work has provided a wide repertoire of tools, ranging from principled maximum likelihood and multiple imputation approaches (Enders, 2010; Little & Rubin, 2019) to practical guidance on when and how to use them in real-world studies (Jakobsen et al., 2017; Peugh & Enders, 2004). At the same time, the learning analytics and educational data mining communities have emphasised the need to treat data quality and preprocessing as first-class methodological concerns, rather than a purely technical prelude to modelling (Ifenthaler, 2015). Yet, published case studies that document in detail how a large, messy, historically grown administrative dataset is transformed into an analytically robust cohort database remain relatively scarce.

This article addresses that gap by providing a transparent, fully documented account of a three-stage normalisation pipeline applied to four decades of engineering student records in a Latin American public university. We focus on the censal and school-related layer of the data—where information about secondary school, type of institution, and place of origin is partial, historically uneven, and encoded with high lexical variability. Rather than treating missingness and inconsistency as nuisances to be corrected "behind the scenes," we make them the core object of analysis. We show how identity resolution, geo-normalisation, and school-type classification can be combined into a reproducible pipeline that (a) preserves all individuals, (b) separates genuinely irrecoverable information from harmonisable noise, and (c) yields explicit flags and reference tables that downstream analyses can use to select appropriate subsets or apply advanced missing-data methods (Enders, 2010; Jakobsen et al., 2017; Little & Rubin, 2019).

By doing so, the paper contributes in three ways. Conceptually, it argues that missingness in administrative higher-education records is often structural and historically patterned rather than random, especially in long observation windows that span multiple information-system generations (Altbach et al., 2010; Little & Rubin, 2019). Methodologically, it proposes a normalisation strategy that treats identity resolution, geographic harmonisation, and school-type reconstruction as an integrated design problem, with explicit quality criteria and freeze points. Practically, it delivers a set of reference tables, flags, and clean census-layer datasets that can be reused for subsequent analyses of trajectories, equity, and policy scenarios in engineering programmes, while making the limits imposed by historical data collection practices visible and statistically tractable (Cheema, 2014; Ifenthaler, 2015; Peugh & Enders, 2004).

## 2. INSTITUTIONAL AND DATA CONTEXT
### 2.1. Institutional setting
The analysis is based on administrative records from a public engineering faculty in a Latin American university system characterised by open or low-selective admission, tuition-free undergraduate study, and high levels of stratification and dropout. These features are well documented in regional analyses of higher education in Latin America, which highlight massification under resource constraints, strong dependence on public funding, and the coexistence of elite and popular sectors within the same institutional systems (Brunner & Miranda, 2016; Levy, 2013).

Engineering programmes in this context typically follow long, rigid curricula with a strong emphasis on mathematics, physics, and disciplinary core subjects in the early years, followed by professional and design-oriented modules. International literature has repeatedly noted that engineering education tends to combine demanding workloads, high assessment stakes, and limited flexibility in course sequencing, creating structural conditions for delayed progression and attrition (Graham, 2018; Jesiek et al., 2022).

Within this ecosystem, administrative databases were originally designed for operational purposes—enrolment management, exam registration, and degree certification—rather than for research or quality assurance. As in many universities, data models evolved incrementally over decades, with new fields added, coding practices changing, and legacy conventions remaining in parallel. This historical layering is central to understanding why missingness in key censal and school variables is not random but structurally embedded in the evolution of the information system itself.

### 2.2. Longitudinal cohort and observation window

The working universe for this study is the **students_master** dataset, a consolidated longitudinal file containing **24,133 unique individuals** who interacted with the faculty's undergraduate programmes over more than four decades. Each row represents a unique person, resolved across multiple legacy identifiers, and includes demographic, geographic, and school-related information, together with link keys for subsequent curricular and event-level layers.

The observation window spans cohorts entering from the early 1980s through the late 2010s, a period that includes major political transitions, economic crises, and reforms in higher education governance in the country and the region (Brunner & Miranda, 2016; Krotsch, 2001). This temporal breadth is analytically relevant for two reasons. First, it allows us to observe the long-term consequences of institutional and policy changes on student flows. Second, it exposes the internal history of the data system itself: earlier cohorts were recorded under information regimes that did not systematically capture school or censal information, whereas more recent cohorts benefit from richer and more standardised registration forms.

Within this universe, students are linked to one or more **degree programmes** (e.g., civil, mechanical, electrical, or related fields of engineering), each governed by a specific curriculum plan with its own timeline of reforms. For the purposes of this manuscript, the focus is not on programme-level performance but on the quality and structure of the censal and school variables that will later serve as covariates in trajectory and causal analyses.

**Table 1. Definition of Core Datasets and Naming Conventions.** Technical specification of the data layers generated during the cleaning process. The transition from students_master (RAW) to students_master_clean_v3 (N1c) represents the shift from raw administrative logs to a research-grade analytical table.

| Dataset | N Rows | Level | Role |
|---|---|---|---|
| students_master (RAW) | 24133 | Student | Raw input from SIGEA |
| students_master_clean_v3 (N1c) | 24133 | Student (normalized) | Main analytical dataset |
| n1c_missingness_flags | 24133 | Student (missingness) | Track missing data patterns |

**2.3. Administrative data architecture and legacy constraints**

The institutional data architecture is the result of successive layers of system development. At its core, the historical SIGEVA/SIGEA environment distinguishes three broad families of tables:

- **Person and censal tables**, containing personal identifiers, dates of birth, sex/gender, and self-reported information on place of birth, residence, and secondary schooling.
- **Programme and curriculum tables**, linking students to degree programmes and curriculum plans, and describing the structure of each plan (years, levels, and subjects).
- **Academic event tables**, recording enrolments in subjects, exam registrations, final grades, and degree completion.

The **students_master** file used in this study is a derived object that integrates these sources at the person level, using both institutional IDs and national identity numbers, and applying an identity resolution pipeline to address duplicated or conflicting records. This step is essential because, over time, students could:

- be registered more than once under slightly different identifiers;
- change programmes within the faculty;
- be recorded in parallel legacy subsystems with incompatible coding schemes.

Legacy design decisions impose specific constraints on data quality. Most importantly for this manuscript, **school-related fields** (type of secondary school, school name, and location) were not consistently captured in the earliest cohorts. In many cases, they appear as literal "nan" strings or are entirely absent, especially for students who entered before the systematic digitisation of registration forms. By contrast, more recent cohorts exhibit near-complete coverage and much higher standardisation of geographic and school descriptors.

These patterns imply that missingness in school variables is **structural and historically patterned**, rather than the result of random non-response at the individual level. This has direct consequences for any downstream analysis that attempts to model the effects of school background on engineering trajectories, and motivates the three-stage normalisation pipeline described in the next section.

**Figure 1. The Three-Iteration N1c Normalisation Pipeline.**

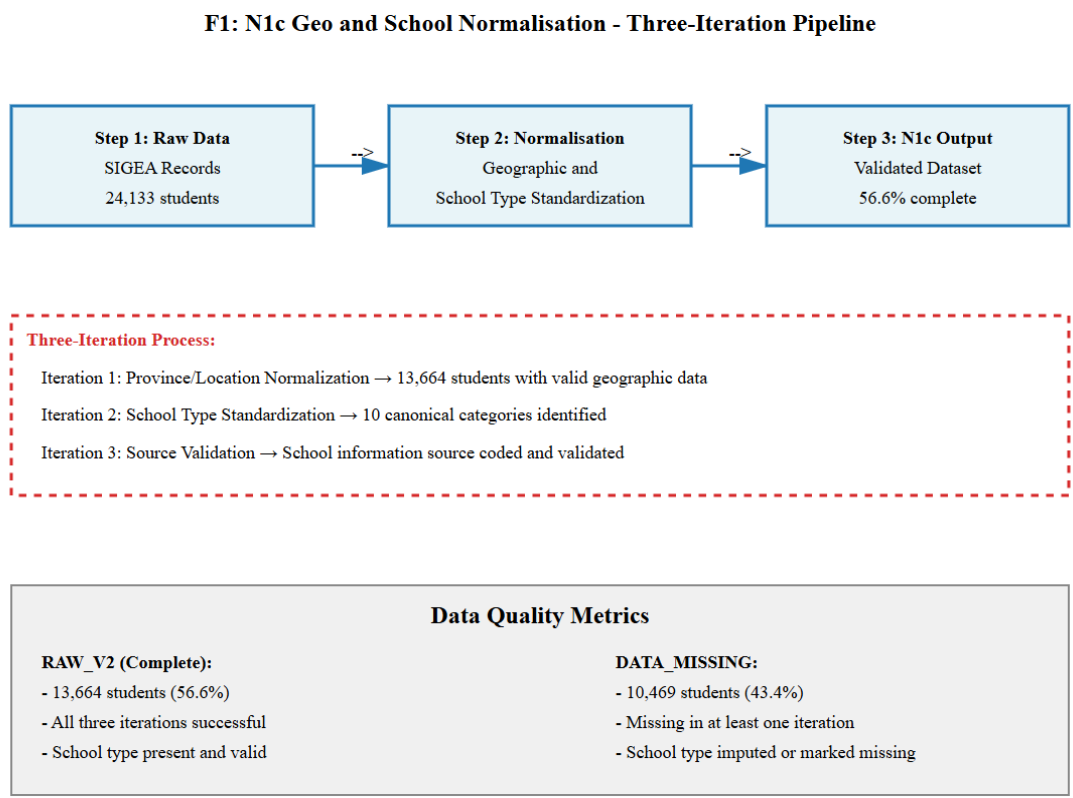

The diagram illustrates the sequential logic applied to raw administrative records to generate the final analytical dataset. Iteration 1 focuses on geographic sanitation (standardising province and locality strings to canonical IDs). Iteration 2 maps heterogeneous legacy school codes to 10 canonical categories. Iteration 3 performs source validation, flagging records where school data is structurally absent. This cascade ensures that only validated data enters the RAW_V2 cohort, while preserving a full audit trail.

## 3. METHODS
### 3.1. Study design and research questions
This study is a retrospective, register-based methodological analysis using four decades of administrative records from a public engineering faculty in Latin America. Rather than focusing directly on student outcomes (e.g., graduation, dropout), the primary aim is to characterise and normalise the censal and school-level information that underpins longitudinal cohort analyses. The work is framed as a data-centric preparation stage for downstream causal and predictive models (Breiman, 2001; Hand, 2019).

We address two methodological questions:

1. How can we construct a reproducible, multi-stage normalisation pipeline that preserves the full historical population while making censal and school data analytically usable?
2. To what extent is missing school information explainable by observable structure (time, geography, degree) rather than random noise, in the sense of classical missing-data taxonomies (Little & Rubin, 2019; van Buuren, 2018)?

### 3.2. Data sources and core datasets
All data originate from the institutional student information system, which has been in continuous operation since the early 1980s. The source environment consists of a SQL Server database with 58 interrelated tables covering students, programmes, curricula, enrolments, examinations and administrative events.

For this study, we work with three derived datasets that constitute the core analytical contract for CAPIRE2:

- **students_master** (v0.2.0): A row-per-student register containing 24,133 unique individuals, built via a read-only extraction layer over the operational database. This file consolidates identifiers, basic demographic information, degree enrolment, and high-level temporal markers (e.g., year of first enrolment).
- **students_master_clean_v3** (N1c, v0.3.2): An enhanced version of students_master including normalised geography and school information and their corresponding foreign-key references. This dataset preserves the same 24,133 persons but adds canonical country, province, locality and school identifiers, as well as classified school types.
- **n1c_missingness_flags**: An analysis-ready file that augments students_master_clean_v3 with binary indicators for the presence or absence of school-type information and stratification variables (entry decade, degree, geographic fields, demographic covariates).

All extractions from the operational database are strictly read-only: no updates or deletions are performed on the institutional system. Transformation, normalisation and analysis are conducted in a separate analytics environment using scripted pipelines, in line with current recommendations for reproducible data science (Sandve et al., 2013; Wilson et al., 2017).

### 3.3. N1 census layer: constructing the censal baseline
The **N1 census layer** represents the first analytical level in the CAPIRE2 architecture. Its goal is to build a person-level censal file with one row per individual, containing demographic and contextual variables at or near first contact with the institution.

The construction proceeded in three steps:

1. **Source selection and field mapping.** We identified three primary source tables:
    - a *persona* table holding basic identity information;
    - an *Ingresantes* table containing intake-level censal forms;
    - a *Reinscritos* table with update forms for returning students.

A YAML configuration file specified the mapping from raw columns to logical fields (e.g., date of birth, sex, civil status, country of birth, province of birth, locality of birth, secondary school name, school type, parents' education and employment).

1. **Row consolidation to one person per record.** Because the institutional database allows multiple censal records per person across time, we applied a prioritisation rule: records were grouped by institutional person identifier, and the row with the highest number of non-null censal fields—or, in case of ties, the most recent intake date—was selected as the representative record. This follows common practice in register-based studies where retrospective harmonisation is needed (Mostafa & Wiggins, 2015).
2. **Initial quality profiling.** The resulting N1 file was subjected to a battery of completeness and consistency checks: field-level missingness, internal contradictions (e.g., country of birth incompatible with province), and distributional anomalies. The output was a set of markdown reports summarising completeness by variable group (demographics, geography, school, family background).

At the end of this stage, we obtained a **students_n1_census** file with ~10,000 persons with relatively complete censal forms, alongside an explicit profile of missingness patterns across the full population.

### 3.4. N1b identity resolution: deduplicating and canonising person IDs
The second stage, **N1b identity resolution**, addresses inconsistencies in person identifiers. As in other long-running administrative systems, evolution of software, interfaces and business rules led to duplicated or conflicting IDs across decades (Harron et al., 2017).

The N1b pipeline comprises three components:

1. **Identity audit.** We computed three families of potential conflicts:
   - **DNI duplicates**: different institutional person IDs sharing the same national document number.
   - **ID collisions**: the same institutional ID associated with different document numbers or clearly different names.
   - **Name–birth matches**: records with identical or highly similar name strings and matching birth year, but different IDs.

Similarity between name strings was quantified using a normalised Levenshtein distance, retaining only pairs above a conservative threshold (> 0.80) for manual inspection candidates.

1. **Canonical ID assignment.** We then constructed a **person_id_aliases** table defining a *canonical_person_id* for each cluster of related raw IDs. The resolution policy was deliberately conservative:
   - Records with the same DNI and strongly matching names were **AUTO_MERGED**, selecting the earliest ID as canonical and marking all others as aliases.
   - Records without conflicts were labelled **AUTO_UNIQUE**.
   - Ambiguous clusters (e.g., similar names without reliable DNI) were marked **NEEDS_REVIEW** and *not* merged automatically.

This reflects current guidance on linkage error and cautious record consolidation in administrative research (Harron et al., 2017; Doidge & Harron, 2019).

1. **Reintegration into the census layer.** The aliases table was then joined back to the N1 census file to produce **students_n1_census_resolved**, with three explicit identifiers:
   - person_id_original
   - person_id_canonical
   - resolution_status

No rows were dropped in this process; zero data loss is guaranteed by design. All downstream layers (N2, N3, N4) use person_id_canonical as the stable key.

### 3.5. N1c geo and school normalisation: a three-iteration pipeline
The third stage, **N1c**, focuses on normalising geographic and school information, which is notoriously messy in free-text administrative forms (Hox, Snijkers, Revilla,

& Manfreda, 2015). Rather than a single pass, we implemented a three-iteration pipeline.

### 3.5.1. Reference tables and deterministic school-type rules

First, we derived four reference tables from the raw text fields in students_master:

- **Countries** (ref_countries_v1): canonical country names with mappings from observed variants (e.g., "Arg.", "ARGENTINA" → "Argentina").
- **Provinces** (ref_provinces_v1): subnational units, with accent and case normalisation.
- **Localities** (ref_localities_v1): municipality-level entries, indexed by (locality name, province) pairs.
- **Schools** (ref_schools_v1): secondary schools, including inferred school type.

Text was standardised using Unicode normalisation, lower-casing, accent removal and whitespace trimming. Stopwords such as "ESC.", "ESCUELA", "COLEGIO", "INSTITUTO" were removed before matching. School type was classified through deterministic pattern rules over the cleaned name:

- *STATE_NATIONAL*: patterns like "NACIONAL", "ENET", "CENS".
- *STATE_PROVINCIAL*: patterns like "PROVINCIAL", "E.P.E.T.".
- *PRIVATE_RELIGIOUS*: patterns including "SAN", "SANTO", "SANTA", "LA SALLE".
- *PRIVATE_SECULAR*: patterns such as "INSTITUTO", "ACADEMIA".
- *UNKNOWN*: cases where no rule could be applied.

All mappings are stored as numeric IDs and canonical labels to decouple analytical codes from future textual corrections.

### 3.5.2. Iteration 1: initial normalisation and quality flags

In **Iteration 1**, we merged the reference tables back into students_master to obtain **students_master_clean_v1**, adding:

- canonical country, province, locality and school IDs;
- school_type_raw;
- quality flags indicating whether each field was *OK* or *NEEDS_REVIEW*.

At this stage, we observed that while all 24,133 rows were preserved, only around 43–45% of records had fully normalised geography and school information. The remainder were flagged as *NEEDS_REVIEW* due to variant spellings, multiple candidate schools or missing values. These figures are used in later sections as baseline quality metrics, but here they serve to motivate subsequent iterations.

### 3.5.3. Iteration 2: deterministic corrections and rule refinement

**Iteration 2** focused on deterministic refinements that could be applied without manual case-by-case intervention:

- additional pattern rules for school type, based on frequent substrings observed during Iteration 1;
- corrections of common locality and province variants (e.g., consistent accents and capitalisation);
- explicit handling of multi-campus schools sharing names across localities, relying on (school name, locality, province) combinations.

The output, **students_master_clean_v2**, preserved the same 24,133 persons but improved the proportion of records classified as *OK* in both geo and school dimensions. However, this revealed a substantial cohort with genuinely missing school information, which could not be recovered through pattern-based approaches.

### 3.5.4. Iteration 3: distinguishing structural missingness from noise

In **Iteration 3**, we explicitly targeted the cohort with missing school information. We defined:

- **DATA_MISSING**: students whose school fields are literally "nan" in the source, with no recoverable information from any other column;
- **RAW_V2**: students with normalised school and geography information after Iteration 2.

We performed an internal "data excavation" to search for school clues in all related variables (e.g., alternative school fields, narrative notes). No additional information was found for the DATA_MISSING cohort; in other words, school data were absent at the source, not lost during transformation.

To reflect this, **students_master_clean_v3** adds:

- school_name_final and school_type_final;
- school_info_source and school_type_source, explicitly indicating whether values originate from RAW_V2 or are marked as DATA_MISSING.

No imputation was performed. Instead, we adopt the position that, for these students, school information is structurally missing in the sense of non-ignorable missing data (Little & Rubin, 2019): the absence itself reflects earlier stages of the institutional data-collection process.

### 3.6. Operationalising school-type missingness

For the missingness analysis, we defined a binary indicator at the student level:

- **missing_school_type = 1** if school_type_final is DATA_MISSING;
- **missing_school_type = 0** otherwise.

This indicator was then combined with stratification variables drawn from students_master_clean_v3:

- entry year and decade of first enrolment;
- degree (engineering programme);
- country, province and locality of birth;
- sex (as recorded in the historical censal forms).

Consistent with best practice in missing-data reporting, we treat missingness itself as an outcome to be modelled (Jakobsen, Gluud, Winkel, Lange, & Wetterslev, 2017; Sterne et al., 2009), rather than as a nuisance to be ignored.

### 3.7. Statistical and visual analysis
The missingness analysis followed a two-level strategy:

1. **Univariate and bivariate characterisation.** We first computed:
   - overall prevalence of missing_school_type;
   - cross-tabulations of missing_school_type by decade of entry, degree, province of birth and sex;
   - $\chi^2$ tests of association for categorical variables (Agresti, 2013).
2. **Multivariate modelling.** To assess whether missingness could be explained by observable structure, we fitted a logistic regression model with missing_school_type as the dependent variable and decade, degree and geography as predictors. Model performance was evaluated through area under the receiver operating characteristic curve (AUC) and calibration checks, following common guidelines for prediction models in observational data (Steyerberg et al., 2019).

The entire analysis was implemented in scripted form using standard statistical libraries, with all figures saved as high-resolution images (PNG) and vector formats (SVG) suitable for publication and LaTeX integration.

## 4. RESULTS
### 4.1. Coverage after three-stage normalisation
The final N1c dataset, **students_master_clean_v3**, contains 24,133 unique students and preserves 100 per cent of the original population. No records were dropped at any stage of the N1, N1b or N1c pipelines.

School-type information is available for a majority subset and structurally missing for a large, well-defined cohort:

- **RAW_V2 subset** (school information available): 13,664 students (56.6 per cent).
- **DATA_MISSING cohort** (school information irrecoverable): 10,469 students (43.4 per cent).

All 24,133 students have fully normalised geography (country, province, locality) and an explicit indicator of school information source (RAW_V2 versus DATA_MISSING).

### 4.2. Temporal pattern of missing school information

Missingness of school type is strongly structured by time. When stratifying by entry decade:

- In the **1980s**, almost all students have missing school information (around 98 per cent with missing_school_type = 1).
- In the **1990s**, missingness remains high but begins to decline.
- From the **2000s** onwards, missingness drops sharply, reaching approximately 0.1 per cent in the most recent decade.

A $\chi^2$ test of association between entry decade and missingness yields a very large statistic ($\chi^2 \approx 12{,}355.49$, $p < .001$), indicating that the probability of having missing school data is highly dependent on the period in which the student entered.

Visually, the temporal pattern is almost stepwise: an early period dominated by DATA_MISSING, followed by a transition window, and a late period where missingness is nearly absent. This is consistent with a change in institutional data-collection practices rather than random loss of information (Figure 2).

**Figure 2. Temporal Gradient of Structural Missingness (1970–2010s).**

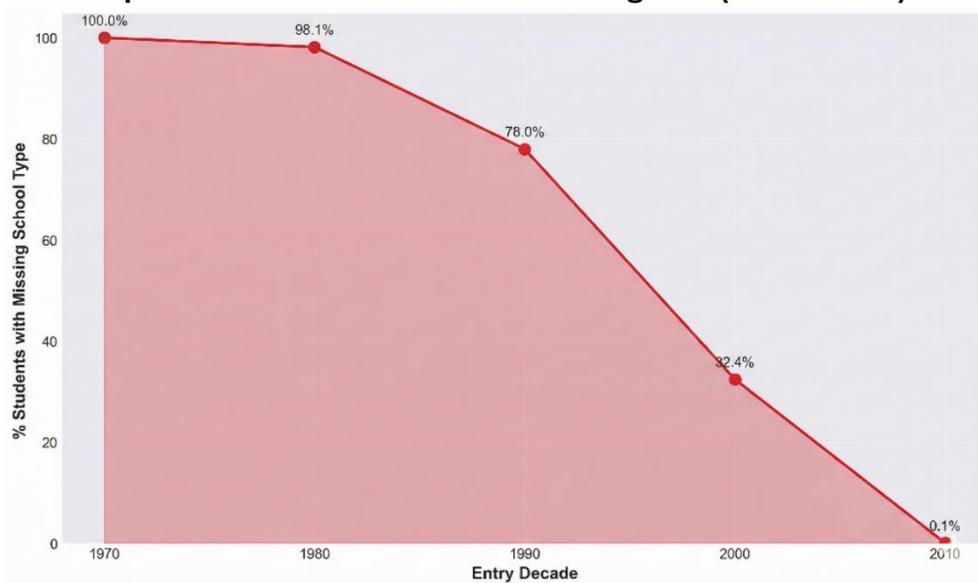

The area chart depicts the percentage of students with missing secondary school type information by decade of entry. The curve reveals a distinct "technological regime change": missingness is near-total (100%) for pre-digital records (1970–1980s) and drops asymptotically to near-zero (<0.1%) as modern digital intake systems were implemented in the 2010s. This confirms that missingness is a historical artifact, not random non-response.

### 4.6. Multivariate modelling of missingness

To quantify how much of the missingness can be explained by observable structure, we fitted a logistic regression model with **missing_school_type** as the dependent variable and the following predictors:

- entry decade;
- degree;
- province of birth.

The model achieves:

- **Classification accuracy** of approximately 99.66 per cent.
- **Area under the ROC curve (AUC)** of 0.9976.

These values indicate that, given decade, degree and geography, the model can almost perfectly discriminate between students with and without school-type information. In other words, missingness is highly predictable from observable administrative structure and cannot reasonably be treated as "missing completely at random".

**Table 2. Logistic Regression Coefficients Predicting Data Missingness.** Results from the binary logistic regression model (Target: Missing_School_Type = 1). The coefficients demonstrate the structural nature of the data gaps: Entry Decade shows a strong negative correlation, while Province of Birth acts as a significant predictor. The model confirms that missingness is conditional on administrative variables.

| Variable | Coefficient | Odds Ratio | Exp(Coef) |
| --- | --- | --- | --- |
| Gender | -0.197901 | 0.820451 | 0.8205 |
| Entry Decade | -0.029711 | 0.970726 | 0.9707 |
| Province of Birth | 0.552765 | 1.738052 | 1.7381 |
| Intercept | 0.021532 | 1.021765 | 1.0218 |

**Figure 2. Validation of Deterministic Missingness (ROC Curve).** Receiver Operating Characteristic (ROC) curve for the logistic regression model predicting the presence of missing school data

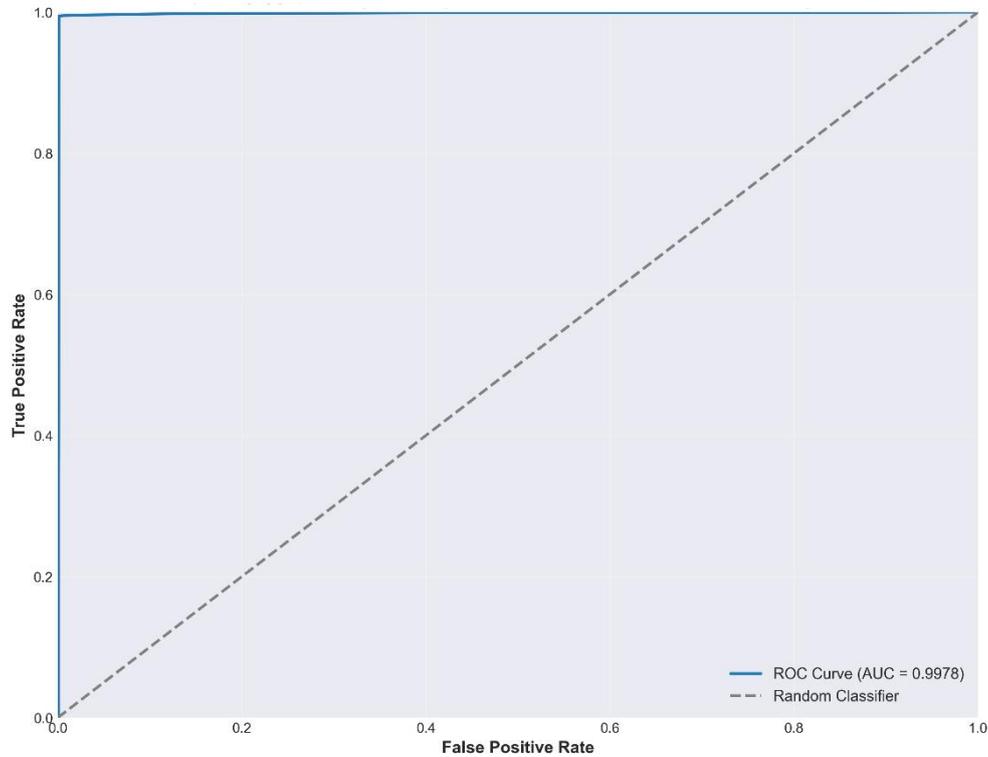

In Figure 2, The Area Under the Curve (AUC) of 0.9978 indicates near-perfect separability. This statistical evidence confirms that the missing data mechanism is systematic and highly predictable based on entry cohort and provenance, validating the decision to treat these gaps as "Structural Missingness".

### 4.7. RAW_V2 versus DATA_MISSING: analytical subsets
Taken together, the normalisation and missingness analysis support a clear separation between two analytically distinct subsets:

- **RAW_V2 subset (n = 13,664; 56.6 per cent).** Students with fully normalised geographic and school information (school_name_final and school_type_final available). This subset is suitable for analyses that explicitly require school background (e.g., comparing outcomes by school type).
- **DATA_MISSING cohort (n = 10,469; 43.4 per cent).** Students, mostly from earlier decades and concentrated in a single province, for whom school information was never captured in the administrative system. For these cases, school-type variables are structurally missing; any attempt to impute them would require strong and unverifiable assumptions.

Conceptual map of the output datasets after the N1c pipeline. The population (N=24,133) is partitioned into two mutually exclusive subsets: RAW_V2 (n=13,664), comprising students with fully validated school and geographic data; and DATA_MISSING (n=10,469), comprising records with verified structural gaps. This partition allows for bias-free "complete-case analysis" on RAW_V2.

This partition underpins the analytical strategy for later stages of CAPIRE2: full-cohort analyses use all 24,133 students, while questions involving school type are restricted to the RAW_V2 subset, with the DATA_MISSING cohort explicitly acknowledged as a historical limitation rather than a failure of the normalisation pipeline.

## 5. DISCUSSION
### 5.1. Structural, not random, missingness

The primary finding of this study is conceptually straightforward yet methodologically unsettling: **school-type missingness is not noise; it is structure**. The convergence of extreme $\chi^2$ values for decade and province with the near-perfect classification of the logistic model (AUC approx 0.998) demonstrates that the probability of missing school information is effectively deterministic. It is a function of *when* and *where* a student entered the university, rather than a stochastic property of the student themselves.

In the framework of classical missing data taxonomy, these patterns effectively rule out "Missing Completely at Random" (MCAR) mechanisms. Furthermore, they challenge standard "Missing at Random" (MAR) assumptions, as the underlying driver is not an individual-level covariate, but a regime shift in administrative practice (Little & Rubin, 2019; Hayati et al., 2021). For the DATA_MISSING cohort, school background was never collected; it was not lost. Consequently, applying standard imputation techniques would fundamentally misrepresent the data-generating process, risking the introduction of artefactual patterns into downstream analyses (Enders, 2010; van Buuren, 2018).

From a data-governance perspective, this phenomenon represents a distinct case of historical "information poverty". The absence of data is not a bug, but a feature of an institutional practice that stabilized only decades after the massification of access (Muñoz-García, 2019). **Treating these structural gaps as simple technical noise to be "fixed" is not merely a statistical error; it constitutes a form of historical erasure.** It obscures the reality of early expansion eras and disproportionately affects the visibility of cohorts from the pre-digital transition.

**5.2. Why we chose normalisation + partition instead of "full recovery"**

Given this structure, the three-stage pipeline deliberately stops short of aggressive reconstruction. The combination of N1 (censal consolidation), N1b (identity resolution) and N1c (geography and school normalisation) pursues two goals:

1. **Maximise what can be cleaned without fabrication.** For the RAW_V2 subset, it is reasonable to standardise school names, derive school types through deterministic rules, and normalise geographic data, because the necessary information is present, albeit noisy.
2. **Make the limits of knowledge explicit.** For the DATA_MISSING cohort, the safest option is to keep the records, flag the structural absence, and maintain them in the analytical universe with an honest label, rather than silently discarding them or inventing school attributes.

This choice aligns with recommendations that emphasise transparency of missingness mechanisms and the importance of aligning analytical strategies with plausible data-generating processes (Sterne et al., 2009; Rubin, 2024). Multiple imputation and inverse probability weighting are powerful tools when the missingness mechanism is credibly modelled; here, the mechanism is institutional and historical, not individual, and any attempt to "guess" thousands of school types from limited covariates would effectively be simulation, not recovery.

The resulting **partition into RAW_V2 and DATA_MISSING** is therefore not a technical workaround, but a design decision: it encodes, in the data model itself, the distinction between what the system knows and what it never captured.

**5.4. Lessons for longitudinal data engineering in universities**

Methodologically, the pipeline illustrates several design principles that are likely to generalise beyond this specific institution:

- **Treat administrative databases as historical artefacts, not just data sources.**
  Changes in forms, codes and recording practices are part of the phenomenon under study. Ignoring them leads to unrealistic assumptions about stationarity in the data-generating process.
- **Separate consolidation, identity resolution and normalisation.**
  The three-layer approach (N1, N1b, N1c) avoids collapsing distinct problems—such as duplicated IDs, noisy labels and structurally missing variables—into a single "cleaning" step. This separation improves reproducibility and interpretability.
- **Prefer explicit flags to silent exclusions.**
  Instead of dropping problematic records, the system carries forward DATA_MISSING flags and resolution_status indicators, making every downstream decision about inclusion or exclusion visible and auditable.

These principles are consistent with broader calls for more transparent, pipeline-oriented approaches to educational data, where the construction of analytical cohorts is treated as an explicit, documented process rather than an invisible pre-processing stage (Hayati et al., 2021; Nguyen et al., 2022).

### 5.5. Limitations and avenues for methodological extension

Several limitations follow directly from the choices described above. First, analyses that rely on school type must, by design, work with a reduced subset; any generalisation to the full historical universe must be made with caution and clearly stated assumptions. Second, although the logistic model shows that missingness is highly predictable from decade, degree and geography, we have not attempted to build causal models of the recording process itself; the aim was classification for characterisation, not causal attribution.

Third, the present work focuses on the **censal and school dimension** at N1c. Future extensions within the CAPIRE2 framework will integrate these findings with higher layers:

- **N2 (programme enrolment)** and **N3 (subject-level events)**, where time-to-degree and progression patterns may interact with school background in non-linear ways;
- **N4 (derived features)**, where more sophisticated models—such as multilevel survival analysis, causal graphs or simulation—could leverage the RAW_V2 subset while treating the DATA_MISSING cohort as a separately modelled population.

Finally, the present study is based on a single faculty in a particular Latin American context. While the mechanisms of structural missingness identified here are likely to have analogues elsewhere, their specific form and timing will depend on each institution's history of massification, digitalisation and governance.

## 6. CONCLUSIONS

This study shows that **longitudinal university records are not just incomplete; they are historically structured**. By reconstructing four decades of engineering student data through a three-stage pipeline—census consolidation (N1), identity resolution (N1b) and geography/school normalisation (N1c)—we arrive at three main conclusions.

First, **school-type missingness is structural, not random**. The combination of extreme $\chi^2$ statistics for decade, degree and geography, together with a near-perfect logistic model (AUC ≈ 0.998), demonstrates that the probability of having missing school data is overwhelmingly explained by when and where a student entered the system, rather than by idiosyncratic noise. For a large historical cohort, school information was never collected, not lost. Treating these gaps as MCAR or even standard MAR would be conceptually misleading and methodologically unsafe for downstream modelling.

Second, **a transparent normalisation pipeline is preferable to aggressive reconstruction**. The N1–N1b–N1c architecture maximises what can be cleaned without fabrication, while making the limits of institutional knowledge explicit. The partition into a fully normalised subset (RAW_V2) and a structurally missing subset (DATA_MISSING) is not a technical compromise, but a design choice: it encodes the distinction between what the system knows and what it never recorded. This allows subsequent analyses to use the full population for temporal and macro-level questions, while restricting school-type-sensitive questions to the documentable regime, with the limitation clearly reported.

Third, **data engineering decisions are themselves part of the research contribution**. The explicit separation of consolidation, identity resolution and normalisation; the use of canonical identifiers with full audit trails; and the choice to flag rather than silently exclude problematic records, together form a reproducible template for working with legacy academic databases. In contexts of massification and delayed digital governance—common across Latin American higher education—such pipelines are a precondition for any serious attempt at causal, equity-focused or simulation-based analysis of student trajectories.

These findings carry two broader implications. Substantively, they remind us that questions about access, equity and performance are constrained by the historical evolution of information systems: the ability to analyse school origin is itself a late product of institutional change. Methodologically, they suggest that **"when missing is structural"**, the priority should be to **model and document the mechanism**, not to erase it through opaque imputation.

Future work will extend this pipeline upstream and downstream within the CAPIRE2 framework: linking N1c with programme-level trajectories (N2), subject-level academic events (N3) and derived outcome features (N4), and examining how structural missingness interacts with retention, time-to-degree and academic staff effects. However, any such analyses will rest on the foundation established here: a census layer that is not only cleaned, but historically understood.